\documentclass[twocolumn,showpacs,preprintnumbers,amsmath,amssymb]{revtex4}

\usepackage{graphicx}
\usepackage{dcolumn}
\usepackage{bm}

\def\be{\begin{equation}}
\def\ee{\end{equation}}
\def\bea{\begin{eqnarray}}
\def\eea{\end{eqnarray}}
\def\by{\left(\begin{array}}
\def\ey{\end{array}\right)}

\def\slash#1{\setbox0=\hbox{$#1$}#1\hskip-\wd0\dimen0=5pt\advance
       \dimen0 by-\ht0\advance\dimen0 by\dp0\lower0.5\dimen0\hbox
         to\wd0{\hss\sl/\/\hss}}

\newcommand{\bl}{{\bf LEFT}}
\newcommand{\br}{{\bf RIGHT}}

\begin{document}

\title{Impact of Nucleon Mass Shift on the Freeze Out Process}

\author{Sven Zschocke}
\affiliation{Section for Theoretical and Computational Physics, and
Bergen Computational Physics Laboratory, 
University of Bergen, 5007 Bergen, Norway}
\affiliation{Forschungszentrum Rossendorf, 01314 Dresden, Germany}
\author{L\'aszlo P\'al Csernai}
\affiliation{Section for Theoretical and Computational Physics, and
Bergen Computational Physics Laboratory, 
University of Bergen, 5007 Bergen, Norway}
\affiliation{MTA-KFKI, Research Institute of Particle and Nuclear Physics,
1525 Budapest 114, Hungary}
\author{Etele Moln\'ar}
\affiliation{Section for Theoretical and Computational Physics, and
Bergen Computational Physics Laboratory, 
University of Bergen, 5007 Bergen, Norway}
\author{Jaakko Manninen}
\affiliation{Section for Theoretical and Computational Physics, and
Bergen Computational Physics Laboratory,
University of Bergen, 5007 Bergen, Norway}
\affiliation{University of Oulu, Department of Physical Sciences, 90571 Oulu,
Finland}
\author{\'Agnes Ny\'iri}
\affiliation{Section for Theoretical and Computational Physics, and
Bergen Computational Physics Laboratory,
University of Bergen, 5007 Bergen, Norway}

\begin{abstract}
The freeze out of a massive nucleon gas through a finite 
layer with time-like normal is studied. The impact of in-medium 
nucleon mass shift on the freeze out process is investigated. 
A considerable modification of the thermodynamical variables   
temperature, flow-velocity, energy density and particle density 
has been found. Due to the nucleon mass shift the freeze out 
particle distribution functions are changed noticeably in comparison 
with evaluations, which use vacuum nucleon mass.  
\end{abstract}

\pacs{}
\maketitle

\section{Introduction}\label{intro}

High-energy nucleus-nucleus collision experiments are mainly 
designed for the search and investigation of the predicted
new state of matter, the Quark-Gluon Plasma (QGP), 
in which quarks and gluons would be set free
from the color confinement observed in normal nuclear matter.
Moreover, heavy-ion reactions 
are expected to exhibit other phenomena of Quantum Chromodynamics (QCD) 
in the hot and dense environment of the collision region, like 
in-medium modifications of almost all hadrons, or the state of 
Color Superconductivity (CSC). In this respect, the 
nucleus-nucleus collision experiments 
provide a unique way to test the validity of present theoretical  
approaches and models of physics of strongly interacting matter.

On the other side, a characteristic and inevitable 
problem of collision experiments is that 
in-medium modifications of hadrons and the 
expected new states of matter (e.g. QGP, CSC)
disappear by the end of the reaction. Accordingly, one can not 
directly measure these properties of the produced hot and dense medium. 
Instead, one has to probe the initial stages of the collision 
indirectly by using theoretical models to reproduce the observed final 
particle spectra. A detailed 
understanding of the different stages of a relativistic  
heavy-ion collision process becomes therefore very compelling.
The scheme of a 
representative relativistic heavy-ion collision process looks as follows:

In the very early stage of nucleus-nucleus collisions, an extremely
hot and dense medium is created in which several hundred or even thousands of 
secondary partons are produced. Due to the high partonic density, local 
(perhaps global) thermal equilibrium is reached very rapidly, 
for instance, at RHIC or LHC incident energies within   
$(0.3 - 0.5)$ fm/c for gluons, and $(0.5 - 1.0)$ fm/c for quarks 
\cite{early_stage_5, early_stage_10, early_stage_15, early_stage_20}. 
It has been proposed that since the heavy quark flavor production
is dominated by the relatively slow gluon-gluon fusion, 
chemical equilibration of the heavy quark flavors (strangeness, charm etc.) 
might stay incomplete during all the collision evolution, so that 
there is a need to implement a strangeness suppression factor 
$\gamma_s$ ~\cite{gammas}.
Nevertheless, chemical equilibration of gluons and light quark flavors
is believed to be reached around $2$ fm/c~\cite{early_stage_25}.

In spite of the nature of the produced medium, large pressure gradient 
perpendicular to the collision axes drives the 
system into rapid expansion and to cool down. In heavy-ion 
collisions, below the critical temperature $T_c \simeq 175$ MeV several 
hundred of hadrons emerge forming a strongly
interacting resonance gas. As the fireball cools down further, 
below the chemical freeze out temperature $T_{ch}$, 
inelastic collisions cease and hadronic abundances become
fixed. This process is usually called the chemical freeze out (cFO). 
Later on, when the hadron gas becomes more dilute, below 
the thermal freeze out temperature $T_{th}$, the elastic 
interactions cease as well. This stage of the collision is usually 
called  kinetic freeze out (kFO). Finally, 
the formed hadrons of the thermal freeze out spectrum propagate 
freely toward the detectors. 
Recently, in \cite{chemical_10} both the
chemical freeze out temperature $T_{ch}$ and thermal freeze out
temperature $T_{th}$ have been determined for
several collision scenarios and baryon densities.
Nonetheless, the sharp distinction between chemical and thermal freeze out 
is an idealization, while in a real collision due to the short time scales 
both processes become mixed with each other.  
Therefore, one sometimes calls it freeze out process without 
further specification between chemical and thermal freeze out which implies    
$T_{ch} \simeq T_{th}$.

Many kind of approaches have been applied for the description of the 
freeze out of strongly interacting matter. Statistical models 
~\cite{chemical_10, becattini1, becattini2, becattini3, cleymans2, 
munzinger1, munzinger2, other_stat_model}
can reproduce well the measured particle multiplicities 
in most of the collision 
experiments done so far. Kinetic models~\cite{kineticmodels_5, kineticmodels_10}
as well as hydrodynamical approaches~\cite{hydromodels_5} have proven to be 
able to describe most of the collective phenomena like the different flow
components in heavy-ion reactions.
However, despite the success in comparison with experiments, the 
in-medium modifications of the hadrons during the freeze out process 
have not been taken into account yet. In most of the former evaluations 
the vacuum parameters of the particles have been implemented. 
To the best of our knowledge the Refs. \cite{Florkowski, Zschiesche}
seem to be the only investigations, where the impact of in-medium 
hadron masses (mesons and baryons) on the particle ratios during the 
chemical freeze out has been studied. A systematic study about the impact of 
in-medium hadron masses on the kinetic freeze out process has not been  
performed yet.

But, implementing the vacuum parameters of hadrons 
for describing the kinetic freeze out 
process is an approximation which may work or may not work, depending on the 
physical system under consideration.  
For instance, both experiments~\cite{pionmassshift_20} and theoretical 
investigations  
~\cite{pionmassshift_5, pionmassshift_10, pionmassshift_15} suggest, that 
pions embedded in a hot and dense medium 
suffer only a small mass change. 
Accordingly, the description of the freeze out process of a purely pion gas 
by means of their vacuum parameters seems to be a reliable approximation. 
On the other side, the mass of kaons can be shifted     
considerably in a hot and dense medium  
\cite{kaonmassshift_5, kaonmassshift_10, kaonmassshift_15}, so 
that taking into account in-medium modifications for the kaon component 
seems to be compelling.

In this work, we study a nucleon gas and  
investigate how strong the impact of an in-medium mass shift of nucleons 
on the freeze out profile is.   
We compare the obtained results with calculations using a vacuum nucleon mass.

The paper is organized as follows: 
The freeze out process within a finite time-like 
layer is considered in Sec. 2. The nucleon mass shift and 
it's implementation 
into the freeze out process are outlined in Sec. 3. 
The results of our study are presented in Sec. 4. Finally,  
in Sec. 5 a summary and outlook are given. Further notations and a brief 
mathematical remark can be found in the Appendix. 

\section{Freeze Out Process within a finite layer}\label{FO}

In this section we are focussing on the last stage of the collision, 
the freeze out process, i.e. we start our investigation 
from the time of collision where  
the expanding and cooling down system reaches a temperature 
$T\le T_c$, where the 
hadronization of the primary parton gas is almost completed.  

The frozen out particles are formed in a layer of finite thickness $L$, 
bounded by two hyper-surfaces:  
the pre-freeze out hyper-surface with 
$T_{{\rm pre}\, {\rm FO}} \simeq T_{c}$ 
and a post-freeze out hyper-surface with 
$T_{{\rm post}\,{\rm FO}} \simeq T_{ch} \simeq T_{th}$. 
These surfaces are defined by the normal $d \sigma_{\mu}$, which in general 
can be a space-like $d \sigma_{\mu} d \sigma^{\mu} < 0$ or 
time-like four vector, $d \sigma_{\mu} d \sigma^{\mu} > 0$.
The diameter $L$ of the layer is of the order of a few 
mean free paths of the particle under consideration. To get an idea 
about the scales we recall that for 
nucleons at ground state saturation density the mean free path 
is about $1$ fm \cite{meanfreepath_5, meanfreepath_10}. 
   
Dynamical models, like hydrodynamical or transport models, 
allow to describe such freeze out processes through the layer. 
In doing so, the hydrodynamical models have certain advantages 
over transport model
calculations. An important one is that, once the equation of state
and initial conditions of the hadronic matter are specified, the
space-time evolution of the system is uniquely determined by the 
hydrodynamic differential equations. 
Especially, this implies that the impact of several equation of state 
may be investigated in a very direct way. Even more, uncertainties
or assumptions made in the underlying kinetic theory of the particles
under consideration are circumvented. In addition, the use of familiar
thermodynamical concepts, like temperature, flow velocity, pressure   
and energy density 
also provide a transparent physical picture of the evolution. Of course,
the basis of applicability of hydrodynamics is the assumption
of local thermal and chemical equilibrium.
In the following we will suppose the validity of these conditions and 
will apply the theory of hydrodynamics for describing the 
thermal freeze out process in a finite space-time layer.

The theoretical description of the kinetic freeze out  
within a hydrodynamical approach has been worked out 
some years ago 
\cite{rethermalization_5, rethermalization_10, 
finitelayer_5, finitelayer_10, finitelayer_15}. 
Very recently, in 
\cite{spacelike_5} and \cite{timelike_5} the formalism 
has been applied to the case of a finite freeze out layer, 
separately both for space-like and time-like normals.  
While the formalism in \cite{spacelike_5, timelike_5} has 
been developed for the general case of  
a massive particle, the calculations have been 
performed for a massless pion gas. Here, we will make  
use of the outlined formalism of Ref. \cite{timelike_5} for a time-like layer  
for the case of a massive nucleon. 
In particular, we will implement the in-medium mass modification 
of nucleons traveling through the freeze out layer. It is not 
necessary to repeat the formalism of \cite{timelike_5} in detail. 
Instead, we shall restrict our explanations on the basic concept and will 
only give the equations relevant for our study. 

Local equilibrium implies that the thermodynamical parameters inside the layer 
become space-time dependent, i.e. we have a space-time dependent temperature 
$T(x)$, flow velocity $v(x)$, energy density $e(x)$ and 
nucleon density $n(x)$. 
For evaluating these functions we need the  basic equations of hydrodynamics,  
\bea
\partial_{\mu} N^{\mu} (x) = 0 \;, \quad {\rm and} \quad 
\partial_{\mu} T^{\mu \nu} (x) = 0\;, 
\label{hydro_5}
\eea
where 
\bea
N^{\mu} (x) = \int \frac{d^3 {\bf k}}{k^0} k^{\mu} f(x, k)
\label{hydro_15}
\eea
is the particle current, and 
\bea
T^{\mu \nu} (x) = \int \frac{d^3 {\bf k}}{k^0} k^{\mu} k^{\nu} f(x, k)  
\label{hydro_10}
\eea
is the energy momentum tensor. Here, 
$x^{\mu} = (t, {\bf r})$ is the four-coordinate and 
$k^{\mu} = (E_k, {\bf k})$ is the four-momentum of the nucleon. 
While the first relation in (\ref{hydro_5}) is 
only valid when the total number of particles is conserved, 
the second relation in (\ref{hydro_5}) is always satisfied and asserts 
the energy and momentum conservation.
The one-particle distribution function $f(x, k)$ is an invariant 
Lorentz scalar, and 
is normalized to the invariant 
number of particles $N$ (in our case the nucleons), 
i.e. $N = \int d^3 {\bf r} \; d^3 {\bf k} \; f (x, k)$. 

While the components of the tensors (\ref{hydro_15}) and (\ref{hydro_10}) 
depend on the Lorentz frame chosen,
two Lorentz invariant scalars can be obtained,
the invariant scalar energy density $e$ and invariant scalar 
particle density $n$:
\bea
e (x) &=& u_{\mu} (x) \, T^{\mu \nu} (x) \, u_{\nu} (x) \;,
\label{hydro_20}
\\
\nonumber\\
n (x) &=& u_{\mu} (x) \, N^{\mu} (x) \;.
\label{hydro_25}
\eea
We notice that these invariant scalars have to be distinguished from the 
non-invariant energy density ${\tilde e} = E/V$ and particle density  
${\tilde n} = N/V$, where $E$ is the non-invariant total energy and 
$V$ is the non-invariant volume of the system.  

While the invariant relations (\ref{hydro_20}) and (\ref{hydro_25}) 
are valid in any Lorentz frame, in 
a concrete evaluation one has to specify the frame in which
the components of the four-current, energy-momentum tensor and 
the (always time-like) four-velocity $u_{\mu}$ are evaluated. 
Any Lorentz frame can be defined by a Lorentz boost in respect 
to the local Rest Frame of the nucleon Gas, RFG, on which the condition 
$u^{\mu}_{\rm RFG} (x) = (1, 0, 0, 0)$ is imposed; obviously, in 
RFG we have ${\tilde e} = e_{\rm RFG} $ and ${\tilde n} = n_{\rm RFG}$. 
However, this condition does not define the RFG uniquely. There are, 
in general, several 
possibilities to define such a rest frame. Here, we will
take Eckart's definition \cite{book_1}, 
which is the most appropriate one for heavy-ion
reactions with high baryon densities. According to this definition
the local Rest Frame is tied to conserved particles, which can be
achieved by equating the unit vector of the particle four-current with the
four-velocity of the particle flow,
\bea
u^{\mu} (x) &=& \frac{N^{\mu} (x)}{\sqrt{N^{\nu} (x) N_{\nu} (x)}}\;.
\label{hydro_30}
\eea
Accordingly, in RFG  there is no particle flow in spatial directions. 
It is straightforward to recognize, that the Lorentz invariant denominator in 
(\ref{hydro_30}) is just the invariant scalar 
particle density of Eq.~(\ref{hydro_25}). 
And, while the components of four-vectors $u^{\mu}$ and $N^{\mu}$ 
depend on the Lorentz frame chosen, the tensor relation (\ref{hydro_30}),  
which connects these frame-dependent components, 
remains valid in any frame.

From the definitions (\ref{hydro_20}), (\ref{hydro_25}) and (\ref{hydro_30})
one obtains the following set of three coupled differential equations, which, 
by means of Eckart's definition, are valid in any Lorentz frame:   
\bea
d e (x) &=& u_{\mu} (x) \, d T^{\mu \nu} (x) \, u_{\nu} (x) 
\nonumber\\
&& + \, 2 \, d u_{\mu} (x) \, T^{\mu \nu} (x) \, u_{\nu}\;,
\label{hydro_35}
\\
\nonumber\\
d n (x) &=& u_{\mu} (x) \, d N^{\mu} (x) \;,
\label{hydro_40}
\\
\nonumber\\
d u^{\mu} (x) &=& \frac{1}{n (x)} \;
\bigg( g^{\mu \nu} - u^{\mu} (x) u^{\nu} (x) \bigg) \; d N_{\nu}\;.
\label{hydro_45}
\eea
Altogether, since there are four unknowns in the problem under consideration, 
namely $T, v, e, n$, an additional constraint is necessary to get a complete 
system of equations, which uniquely determines these four unknowns.  
That constraint is provided by the Equation of State (EoS)
for the nucleon gas \cite{EOS_5, EOS_10, EOS_15}, which is assumed to be valid
in any space-time point of the reaction zone after hadronization,
\bea
e (x) &=& n (x) \bigg[M_N (n(x) ,T(x)) \, - \,  E_0 
\nonumber\\
&& + \frac{K}{18} \,
\left(\frac{n (x)}{n_0} - 1 \right)^2 \, + \, \frac{3}{2} \; T (x)  \bigg] \;.
\label{eos_5}
\eea
The term $E_0 = 16$ MeV accounts for the nuclear binding energy among the
nucleons, the term proportional to the compressibility constant
$K = 9 \, (\partial p / \partial n)_{n = n_0} \simeq 235$ MeV 
accounts for the dependence of compressibility on density.
Since we are aiming at investigating the impact of in-medium nucleon mass 
shift on the freeze out process, we have 
already implemented a density and temperature dependent 
nucleon pole mass in (\ref{eos_5}). 
The given EoS (\ref{eos_5}) is a generalization of EoS for the 
ideal nucleon gas, which is valid in the rest frame, i.e. $e_{\rm RFG}  
= n_{\rm RFG} [ M_N + 3/2\, T_{\rm RFG}]$. 
There are other generalizations for the nucleonic EoS 
\cite{book_1}. However,
we have checked that in the energy and temperature
region we are working here, the results obtained are insensitive on
the specific choice of nucleonic EoS.
The EoS (\ref{eos_5}) is used to determine the temperature $T (x)$ of
the interacting component of 
the nucleon gas during the freeze out process. Accordingly, the
four equations (\ref{hydro_35}), (\ref{hydro_40}), (\ref{hydro_45}) and
(\ref{eos_5}) represent a closed set for evaluating the four unknowns
$T, v, e, n$ of the one-particle system.

Now we will turn to the explicit evaluation of components 
for the energy-momentum tensor and nucleon four-current. 
In line with Ref. \cite{timelike_5} we will perform all evaluations
in the Lorentz Rest Frame of the freeze out Front, RFF, 
so that in our study all tensor components in 
Eqs.~(\ref{hydro_35}) - (\ref{hydro_45}) 
can be labeled by RFF. The Lorentz frame RFF 
is defined as follows:
 
At a given instant in the space-time the
expanding hot and dense hadronic system reaches
a certain freeze out temperature $T_{{\rm post}\,{\rm FO}}$, 
where all constituents of the system are assumed to get 
frozen out, i.e. all hadrons do not interact anymore.           
In an arbitrary but fixed direction ${\bf e}_x = 
{\bf r}_{\rm T}/|{\bf r}_{\rm T}|$ transverse 
to the beam the              
Rest Frame of the gas RFG moves with a velocity ${\bf v}_{\rm T}$ 
relative to the 
freeze out front RFF. Then, by means of a Lorentz transformation 
the particle four-velocity in RFF becomes 
$u^{\mu}_{\rm RFF} = \gamma (1, v, 0, 0)$ where $v = {\rm sign} 
({\bf v}_{\rm T}) 
|{\bf v}_{\rm T}|$ and 
$\gamma = 1 / \sqrt{1 - v^2}$. 
The velocity $v$ is called
flow-velocity and, in general, can be positive, negative or even zero. 

Furthermore, as the system expands and cools down the number of interacting
particles decreases up to the post freeze out surface of the finite layer, 
where by definition the density of interacting particles vanishes.
Accordingly, 
the thermal freeze out process inside the layer can be described by 
decomposing 
the particle distribution function into two components of the matter, an 
interacting part $f_i$ and a non-interacting free part $f_f$, thus
\bea
f (x, k) &=& f_i (x, k) \; + \; f_f (x, k)\;.
\label{kinetic_5}
\eea
According to Eq.~(\ref{kinetic_5}) and by means of (\ref{hydro_15}), 
(\ref{hydro_25}), (\ref{hydro_30}) we have an interacting and a 
non-interacting particle density, 
\bea
n_i (x) &=& \sqrt{N_{\nu\;i} (x)\, N^{\nu}_i (x)}\;,
\nonumber\\
n_f (x) &=& \sqrt{N_{\nu\;f} (x)\, N^{\nu}_f (x)}\;,
\label{kinetic_6}
\eea
with $n = n_f + n_i$. 
On the pre-freeze out we assume to have thermal equilibrium, i.e. 
we have a J\"uttner distribution for $f_i$ as starting one-particle 
distribution function, while by definition $f_f$ 
is zero on the pre-freeze out 
hyper-surface.  
The space-time evolution of the interacting and non-interacting components 
inside the layer is governed by the following differential equations 
\cite{timelike_5}:
\bea
\partial_t \, f_i &=& - \frac{1}{\tau} \left( \frac{L}{L - t} \right) 
\left(\frac{k^{\mu} \, d \sigma_{\mu}}{k_{\mu} \, u^{\mu}} \right) f_i 
\nonumber\\
&& + \frac{1}{\tau_0} [f_{eq} (t) - f_i] \;,
\label{kinetic_10}
\\
\nonumber\\
\partial_t \, f_f &=& + \frac{1}{\tau} \left( \frac{L}{L - t} \right) 
\left(\frac{k^{\mu} \, d \sigma_{\mu}}{k_{\mu} \, u^{\mu}} \right) f_i \;, 
\label{kinetic_15}
\eea
with the time $\tau$ between collisions. 
The J\"uttner distribution is given as \cite{Juttner} 
\bea
f_{eq} (t) &=& \frac{1}{(2 \pi \hbar)^3} 
{\rm e}^{(\mu - k^{\mu} u_{\mu})/T} \;, 
\label{Juttner_5}
\eea
with the chemical potential $\mu$; for the interacting component 
it is determined by Eq.~(\ref{chemical_potential}) given below.  
The second term in (\ref{kinetic_10}) is the re-thermalization term  
\cite{rethermalization_5, rethermalization_10, finitelayer_10, 
finitelayer_15, timelike_5} which describes how fast the interacting 
component approaches the J\"uttner distribution within a relaxation time 
$\tau_0$. Here, we will use the immediate re-thermalization limit 
$\tau_0 \rightarrow 0$, which implies $f_i \rightarrow f_{eq}$ 
faster than $\tau_0 \rightarrow 0$, 
i.e. local equilibrium at all times during the 
freeze out in the following way:
 
First, the layer is subdivided into small intervals.  
Then we calculate the changes $d T^{\mu\nu}$ and $d N^\mu$  
based on their kinetic definitions (\ref{hydro_15}) and (\ref{hydro_10}), 
respectively, with the freeze out distribution $f_i$.   
At the beginning of a time step 
this is considered to be a flux coming from a J\"uttner distribution, 
and continues during the length of the whole time step according 
to the kinetic differential equation (\ref{kinetic_10}) 
(without the re-thermalization term).
Then the remaining distribution is not of J\"uttner type anymore.
Nevertheless, the loss $d T$  and $d N$ are calculated based on the initial
J\"uttner and the escape probability.  When we are at the
end of the time step of such a small interval of the layer, 
we have a change in all thermodynamical variables $T, v, e, n$. 
With $\tau_0 \rightarrow 0$  
we assume an immediate re-thermalization of $T^{\mu\nu}$ and $N^{\mu}$, i.e.  
we define a new J\"uttner distribution with the new values for 
$T, v, n$ at the
end of the time step.  At the next time step we use this new J\"uttner
distribution to calculate the changes of it in the next small 
time interval, and so on.
Accordingly, the last term in (\ref{hydro_35}) vanishes, as it can be seen as
follows. Since at the beginning of a time step we take a J\"uttner distribution
according to the immediate re-thermalization limit, the second term of 
(\ref{hydro_35}) is zero (see Appendix). 
Then, during a time step the energy momentum tensor 
(\ref{hydro_10}) of a J\"uttner distribution 
is changed by an amount of $d T^{\mu \nu} \sim d t$,
governed by Eq.~(\ref{kinetic_10}). That means
the second term in (\ref{hydro_35}) is
of order ${\cal O} (d u \; d t)$, i.e. of second order in the
differentials, so that the second term in (\ref{hydro_35}) has to be neglected.
For more details about the relations (\ref{kinetic_10}) 
and (\ref{kinetic_15}) and about the re-thermalization limit 
we refer the interested reader to Refs. \cite{rethermalization_5, 
rethermalization_10, finitelayer_15, timelike_5}. 

By means of the microscopic definitions (\ref{hydro_15}) and (\ref{hydro_10})
one obtains for the change of the four-current and  energy momentum tensor 
the following general expressions for the interacting component ,
\bea
d N^{\mu}_i &=& dt \int \frac{d^3 \, {\bf k}}{k^0} k^{\mu} \bigg[ \partial_t 
\, f_i \bigg]
\;, 
\label{change_5}\\
\nonumber\\
d T^{\mu \nu}_i &=& dt \int \frac{d^3 \, {\bf k}}{k^0} k^{\mu} k^{\nu} 
\bigg[ \partial_t \, f_i \bigg]\;.
\eea 
Since we are mainly interested on the freeze out of the  
interacting nucleons we will write down the interacting 
component of these tensors and drop the index $i$ in the following. 
The non-interacting components can be deduced from them 
by changing the sign in front. 
We will write down these expressions explicitly for the change of 
$d N$ and $d T$ 
as given in Ref.\cite{timelike_5} for the RFF, which, as previously  
mentioned, are related to the RFG by a Lorentz boost: 
\begin{widetext}
\bea
\frac{d N^{0} (t, v, T, M_N, n)}{d t} &=& \frac{1} {\tau}
\frac{L}{L - t} \frac{n}{4} \left(
G^{-}_1 (M_N, v, T) - G^{+}_1 (M_N, v, T) \right)  \;,
\label{freezeout_5}
\\
\nonumber\\
\frac{d N^{x} (t, v, T, M_N, n)}{d t} &=& \frac{1}{v}
\frac{d N^{0} (t, v, T, M_N, n)}{d t} + \frac{1} {\tau} \frac{L}{L - t}
\frac{n}{4} \left( \frac{4 a K_1 (a) }{v} + \frac{2 a^2 K_0 (a) }{v} \right)\;,
\label{freezeout_10}
\\
\nonumber\\
\frac{d T^{0 0} (t, v, T, M_N, n)}{d t} &=& \frac{1} {\tau}
\frac{L}{L - t} \frac{n T}{4} \frac{1}{\gamma \, v}
\left(G^{-}_2 (M_N, v, T) - G^{+}_2 (M_N, v, T) \right) \;,
\label{freezeout_15}
\\
\nonumber\\
\frac{d T^{0 x} (t, v, T, M_N, n)}{d t} &=& \frac{1}{v}
\frac{d T^{0 0} (t, v, T, M_N, n)}{d t} 
\nonumber\\
\nonumber\\
&& + \frac{1}{\tau} \frac{L}{L - t}
\frac{n T}{2} \frac{b^2}{v} \left( (3 + v^2) K_2 (a) + a \, K_1 (a) \right) \;,
\label{freezeout_20}
\\
\nonumber\\
\frac{d T^{x x} (t, v, T, M_N, n)}{d t} &=& \frac{1}{v}
\frac{d T^{0 x} (t, v, T, M_N, n)}{d t}
\nonumber\\
\nonumber\\
&& - \frac{T}{\gamma\,v}  \left(\frac{d N^{x} (t, v, T, M_N, n)}{d t}
- \frac{1}{v} \frac{d N^{0} (t, v, T, M_N, n)}{d t} \right)
\nonumber\\
\nonumber\\
&& + \frac{1}{\tau} \frac{L}{L - t}
\frac{n T}{2} a \, b
\left( \frac{1}{v^2} (1 + 3 v^2) K_2 (a) + b \, K_1 (a) \right)\;.
\label{freezeout_25}
\eea
\end{widetext}
Here, $a = M_N/T$ and $b = \gamma \, a$. The functions $G_n^{\pm}$ 
and $K_n$ are defined in the Appendix.  
The set of equations (\ref{hydro_35}) - (\ref{eos_5})  
and (\ref{freezeout_5}) - (\ref{freezeout_25}) allow us to 
evaluate the basic thermodynamical function $T(x), v(x), e(x)$ 
and $n(x)$ during the freeze out process for a particle with a 
constant mass $M_N$. However, as mentioned in the Introduction we 
are aiming at an implementation of in-medium mass shift to look for it's 
impact on the freeze out process. Therefore, 
we will first evaluate the equation with the vacuum nucleon 
pole mass $M_N(0)$, and afterwards  replace it by  
a density and temperature dependent 
nucleon pole mass $M_N (n, T)$.  

\section{Nucleon Mass Shift}\label{mass_shift}

During the freeze out process, the temperature and particle densities are
presumably close to the deconfinement phase transition critical values 
\cite{chemical_10}. Therefore, the in-medium values of masses, decay widths, 
coupling constants,  
and all other physical quantities characterizing the particles under
consideration have to be taken into account.
In our study we examine a purely nucleon gas, and
consider the in-medium mass modification
of nucleons located in a hot and dense nuclear enviroment.

We start with a brief reconsideration of the nucleon mass in vacuum. 
The nucleon derives it's vacuum mass, $M_N(0) = 939$ MeV, from the 
quark-gluon interaction of it's underlying substructure, consisting 
of valence quarks, sea quarks and gluons. However, 
although there has been considerable success 
in reproducing the vacuum mass of nucleons on the basis of their
microscopic quark and gluon substructure 
(lattice evaluations, \cite{lattice_mass}), 
a rigorous use of fundamental theory of QCD 
in this respect is not yet in reach. 
Therefore, our understanding of the nucleon's 
mass structure comes mostly from models. 
From a hadronic field theoretical point of view the nucleon mass $M_N(0)$ 
can be defined as the pole mass 
of the nucleon propagator in vacuum, 
\bea
\Pi_N (k) &=& i \, \int d^4 x \; {\rm e}^{i k x} \;
\langle 0 | {\rm T} \hat{\Psi}_N (x) \hat{\overline \Psi}_N (0) | 0 \rangle  
\nonumber\\
&=& \frac{1}{\gamma_{\mu} k^{\mu} - \stackrel{\rm o}{M}_N - \Sigma_N (k)
+ i \epsilon}\;,
\label{massshift_2}
\eea
where {\rm T} is the Dirac time-ordering, 
$\hat{\Psi}_N$ is the nucleon field operator, 
$\hat{\overline \Psi}_N = \hat{\Psi}_N^{\dagger} \gamma_0$, $\gamma_{\mu}$ 
are the Dirac matrices,  
and $\Sigma_N (k)$ is the nucleon self energy in vacuum. 
The parameter $\stackrel{\rm o}{M}_N$ is called bare nucleon mass, i.e. the 
mass parameter entering the Lagrangian which describes the interaction 
between the nucleons and other hadrons (e.g. nucleon-pion interaction).  
In general, the mass parameter $\stackrel{\rm o}{M}_N$ has to be distinguished 
from the vacuum pole mass of nucleon $M_N (0) = 939$ MeV, defined by
\bea
M_N (0) = \stackrel{\rm o}{M}_N + {\rm Re} \Sigma_N (\gamma_{\mu} 
k^{\mu} = M_N(0))\;.
\label{polemass_5}
\eea
As mentioned, there are several models  
which allow to calculate the pole mass from a QCD based microscopic  
point of view.  
Among them is the extension of QCD sum rule approach 
\cite{Shifman} to the case of baryons \cite{ioffe},  
which provides an interlooking between the nucleon pole mass  
and QCD based quantities, so called QCD condensates.   
Within the QCD sum rule approach, the nucleon field operator $\hat{\Psi}_N$ 
in (\ref{massshift_2})
is expressed by an interpolating field $\hat{\eta}_N$ \cite{ioffe},  
which is made of up and down quark field operators, and  
which has the quantum numbers of a nucleon (charge, spin, isospin, parity). 
In this line, in \cite{ioffe} the so called 
Ioffe formula for the nucleon pole mass in vacuum has been obtained,
\bea
M_N (0) &=& - \frac{8 \pi^2}{M^2} \; \langle 0 | {\overline q} q | 0 \rangle\;, 
\label{nucleonmass_1}
\eea 
providing a link between the pole mass and 
the chiral condensate, $\langle 0 | {\overline q} q | 0 \rangle 
= - (0.250 \;{\rm GeV})^3$;  
$M \simeq 1.15\,{\rm GeV}$ is the Borel mass parameter determined 
by stability constraint of the nucleon sum rule approach \cite{ioffe}.

Now we will turn to the in-medium nucleon pole mass $M_N(n,T)$, 
which is the very  
characteristics which enters the EoS (\ref{eos_5}), \cite{drukarev}. 
In general, a nucleon propagating in a hot and dense hadronic enviroment 
can be regarded as a quasi particle, described by the in-medium 
nucleon correlator  
\bea
&& \Pi_N (k, n, T) 
\nonumber\\
&& = i \, \int d^4 x \; {\rm e}^{i k x} \;
\langle \Omega | {\rm T} \hat{\Psi}_N (x) \hat{\overline \Psi}_N (0) 
| \Omega \rangle \;,
\nonumber\\
&& = \frac{1}{\gamma_{\mu} k^{\mu} - \stackrel{\rm o}{M}_N - \Sigma_N (k, n, T)
+ i \epsilon}\;.
\label{nucleonmass_2}
\eea
The nucleon self energy $\Sigma_N (k,n,T)$ in medium depends on density 
and temperature of the surrounding hadronic medium 
inside the freeze out layer; the hadronic medium is     
described by the state $| \Omega \rangle$.  
In generalization of Eq.~(\ref{polemass_5}) the in-medium nucleon pole mass 
is defined by  
\bea
&& M_N (n,T) 
\nonumber\\
&& = \stackrel{\rm o}{M}_N + {\rm Re} \Sigma_N (\gamma_{\mu}
k^{\mu} = M_N (n,T), n, T)\;.
\nonumber\\
\label{polemass_10}
\eea
From this point of view it becomes obvious that the pole mass will be    
modified in a hot and dense hadronic matter, simply due to the fact
that the self energy of a nucleon in medium will be different from 
a nucleon in vacuum. 

In general, the particle pole mass in (\ref{polemass_10}) for a nucleon 
at rest (RFG), embedded in a hot and dense hadronic medium,     
is given by \cite{nucleonsumrule_10, pole_mass1}  
\bea
&& M_N (n, T) 
\nonumber\\
&& = M_N (0) + {\rm Re} \Sigma_S (n, T) + 
{\rm Re} \Sigma_V (n, T)\;, 
\nonumber\\
\label{massshift_3}
\eea
with the attractive scalar part (${\rm Re}\; \Sigma_S < 0$) and  
the repulsive vector part (${\rm Re}\; \Sigma_V > 0$) of nucleon self energy 
in medium.  
It is a result of several theoretical models applied so far, 
that the individual contributions of scalar and vector self energy  
are large, but they are canceled by each other to a large extent; 
typical values at saturation density are ${\rm Re}\; \Sigma_S = - 400$ MeV,
${\rm Re}\;\Sigma_V = + 300$ MeV 
\cite{nucleonsumrule_10, pole_mass1, values1}.  
In particular, several theoretical approaches predict a mass dropping of
the nucleon pole mass in a hadronic enviroment of about  
$M_N(n_0, 0) - M_N(0) \simeq - (80 \pm 20)$ MeV at 
groundstate nuclear saturation density $n_0 = 0.17\, {\rm fm}^{-3}$ and 
at vanishing temperature. 
Here, we will take the QCD sum rule results for a nucleon 
in matter, given by \cite{nucleonsumrule_10, pole_mass1} 
\bea
{\rm Re} \Sigma_S (n, T) &=& + M_N (0) 
\left( \frac{\langle \Omega | {\overline q} q | \Omega \rangle} 
{\langle 0 | {\overline q} q | 0 \rangle} - 1 \right) , 
\label{massshift_4}
\\
\nonumber\\
{\rm Re} \Sigma_V (n, T) &=& - \frac{8}{3} M_N (0) 
\frac{\langle \Omega | q^{\dagger} q | \Omega \rangle}
{\langle 0 | {\overline q} q | 0 \rangle} \,,
\label{massshift_5}
\eea
where we have accounted for the  
lowest mass dimension condensates only; gluon condensate and higher mass 
dimension condensates give only small corrections due to large cancellations 
between their individual contributions. 
We recall that the part $M_N^{*} \equiv M_N(0) + \, {\rm Re} \Sigma_S$ of 
(\ref{massshift_3}) resembles the terminology "effective mass" 
used in the Walecka model \cite{Walecka}, the Skyrme model 
\cite{nucleonmassshift_10}, and which has also been evaluated   
by means of a mean field approach in \cite{nucleonmassshift_1}.
For a more detailed clarifying of the terminology "effective mass",  
often used in different meaning, we refer to Ref. \cite{effective_mass}, 
where $M_N^{*}$ is called Dirac mass.  

The chiral condensate at finite temperature and density, 
$\langle \Omega | {\overline q} q | \Omega \rangle$,  
has been evaluated within the Nambu-Jona-Lasinio (NJL) model in 
Ref. \cite{change_condensate1}. Later,   
in Ref. \cite{change_condensate2} the in-medium chiral condensate 
has been evaluated at finite densities and temperatures by means of a 
pion-nucleon gas, finding a good agreement with 
the results of Ref. \cite{change_condensate1}. 
Here, the condensates (\ref{massshift_4}) and (\ref{massshift_5}) have to be 
evaluated for a purely nucleon gas to be consistent within the whole approach
presented.
According to Eqs.~(\ref{kinetic_5}) and (\ref{kinetic_6}) there are 
two components inside the finite layer: 
an interacting component with density $n_i$ and a non-interacting  
component with density $n_f$. For evaluating the condensates 
(\ref{massshift_4}) and (\ref{massshift_5}) 
we approximate the interacting component 
by a Fermi gas with chemical potential $\mu_i$ and temperature $T$. 
On the other side, the temperature for the non-interacting component 
becomes ill-defined. Nonetheless, a relevant physical parameter for describing 
the non-interacting component remains the density $n_f$. 
Accordingly, the condensates in one-particle approximation 
are given as follows \cite{change_condensate2}:
\bea
\langle \Omega | {\overline q} q | \Omega \rangle &=& 
\langle 0 | {\overline q} q | 0 \rangle 
\nonumber\\
&& + 4 \int \frac{d^3 {\bf k}}{(2 \pi)^3} 
\frac{1}{2 E_k} N_F 
\langle N ({\bf k}) | {\overline q} q | N ({\bf k}) \rangle 
\nonumber\\
&& + \frac{n_f}{2 M_N(0)} \langle N ({\bf k}) | {\overline q} q | N ({\bf k}) 
\rangle \;,  
\label{condensates_5}
\\
\nonumber\\
\langle \Omega | q^{\dagger} q | \Omega \rangle &=& 
4 \int \frac{d^3 {\bf k}}{(2 \pi)^3} 
\frac{1}{2 E_k} N_F 
\langle N ({\bf k}) | q^{\dagger} q | N ({\bf k}) \rangle 
\nonumber\\ 
&& + \frac{n_f}{2 M_N(0)} \langle N ({\bf k}) | 
q^{\dagger} q | N ({\bf k}) \rangle \;.
\label{condensates_6}
\eea
where $N_F = [{\rm e}^{(E_k - \mu_i)/T}+1]^{-1}$ is the Fermi distribution, 
and the nucleon energy is $E_k = \sqrt{M_N(0)^2 + {\bf k}^2}$.  
Note that $\langle 0 |q^{\dagger} q | 0 \rangle = 0$.
Here, the relativistic normalization $\langle N ({\bf k}_1) |
N ({\bf k}_2) \rangle = 2 E_{k_1} (2 \pi)^3 \delta^{(3)}
({\bf k}_1 - {\bf k}_2)$ is used.
In Eqs.~(\ref{condensates_5}) and (\ref{condensates_6}) the
spin (up, down) and isospin (proton, neutron) degeneracy of nucleon states
has been taken into account by the factor $4$ in front of the 
momentum integrals.
The chemical potential for the interacting component can be
evaluated via
\bea
n_i &=& 4 \int \frac{d^3 {\bf k}}{(2 \pi)^3}
\frac{1}{{\rm e}^{(E_k - \mu_i)/T}
+ 1}\;.
\label{chemical_potential}
\eea
\begin{center}
\begin{figure}[!ht]
\includegraphics[angle=0,scale=0.7]{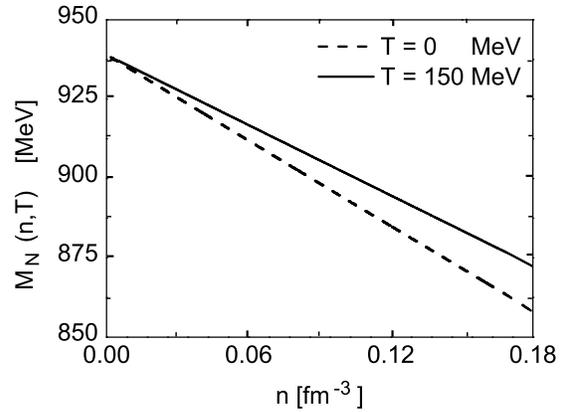}
\caption{
Effective in-medium nucleon pole mass $M_N (n,T)$ according to 
Eq.~(\ref{massshift_3}) 
(for more details see main text).\label{fig:mass}}
\end{figure}
\end{center}
The condensates in Fermi gas approximation are given by \cite{sigmaterm_5}
\bea
\langle N ({\bf k}) | {\overline q} q | N ({\bf k}) \rangle
&=& \frac{M_N (0) \sigma_N}{m_q} \;, 
\label{condensates1}
\\
\nonumber\\
\langle N ({\bf k}) | q^{\dagger} q | N ({\bf k}) \rangle
&=& 3 M_N (0) \;.
\label{condensates2}
\eea
The nucleon sigma term is $\sigma_N \simeq 50$ MeV 
\cite{nucleonsumrule_5}, 
and $m_q \simeq 5$ MeV 
is the averaged current quark mass 
of the up and down quark flavor \cite{PDB, GellMann_Oakes_Renner}.  
Inserting these parameters into (\ref{massshift_4}) and 
(\ref{massshift_5}) we obtain 
${\rm Re} \; \Sigma_S = - 390$ MeV and ${\rm Re}\;\Sigma_V = + 315$ MeV at 
ground state saturation density $n_0$. 
The Eqs.~(\ref{massshift_3}) - (\ref{condensates2})  
summarize our propositions made for obtaining the  
in-medium nucleon pole mass $M_N (n, T)$ which enters the EoS (\ref{eos_5}).
Fig.~(\ref{fig:mass}) shows the dropping of the in-medium nucleon pole mass.
The slight increase of the in-medium nucleon pole mass with temperature is 
an artifact of the purely nucleon gas approximation. That means,  
an implementation of pions in Eqs.~(\ref{condensates_5}) and 
(\ref{condensates_6}) which govern the mass relation (\ref{massshift_3}),  
would cause a temperature decreasing of these condensates 
\cite{change_condensate2} and then of the in-medium nucleon pole mass. 
Here, in a baryon dominated system this artificial increase of $M_N (n,T)$ 
with temperature is superposed by the much stronger down shift 
of the pole mass with nucleon density. 

\section{Results and Discussion}\label{results_discussions}

In this section we represent and discuss the results of the coupled 
set of  differential 
equations (\ref{hydro_35}) - (\ref{hydro_45}) in combination with  
the EoS (\ref{eos_5}) and the 
in-medium nucleon mass shift relations  
(\ref{massshift_3}) - (\ref{condensates2}). 
The differential equations have been solved with the 
Runge-Kutta method \cite{Runge_Kutta1,Runge_Kutta2,Runge_Kutta3} 
on the IBM 1300 cluster at Bergen Center for Computational Science 
(BCCS). For all of the calculations, we have taken 
$T_{{\rm pre}\;{\rm FO}} = 150$ MeV, 
$n_{{\rm pre}\;{\rm FO}} = 1.5 \, n_0 \; 
({\rm corresponding}\; {\rm to}\; 
\mu_{{\rm pre}\;{\rm FO}} \simeq 615 \;{\rm MeV})$ 
and $v_{{\rm pre}\;{\rm FO}} = 0.5 \,c$
as starting values on the pre freeze out hypersurface.
These values are, for instance, in line with typical parameters 
which have been reached within the Alternating-Gradient Synchrotron 
(AGS) at {\it Brookhaven National Laboratory} (BNL) in Brookhaven/USA, 
cf. \cite{BNL}. 
Higher baryonic densities can be reached within the  
Schwer-Ionen-Synchrotron (SIS) at 
{\it Gesellschaft f\"ur Schwerionenforschung} (GSI) 
in Darmstadt/Germany, cf. \cite{SIS}.  
Note that $T_{{\rm pre}\;{\rm FO}}$ and $n_{{\rm pre}\;{\rm FO}}$ 
are pre freeze out values and, 
therefore, they are larger than typical post freeze out values 
given, for instance, in Ref. \cite{chemical_10}. 

\begin{center}
\begin{figure}[!ht]
\includegraphics[angle=0,scale=0.7]{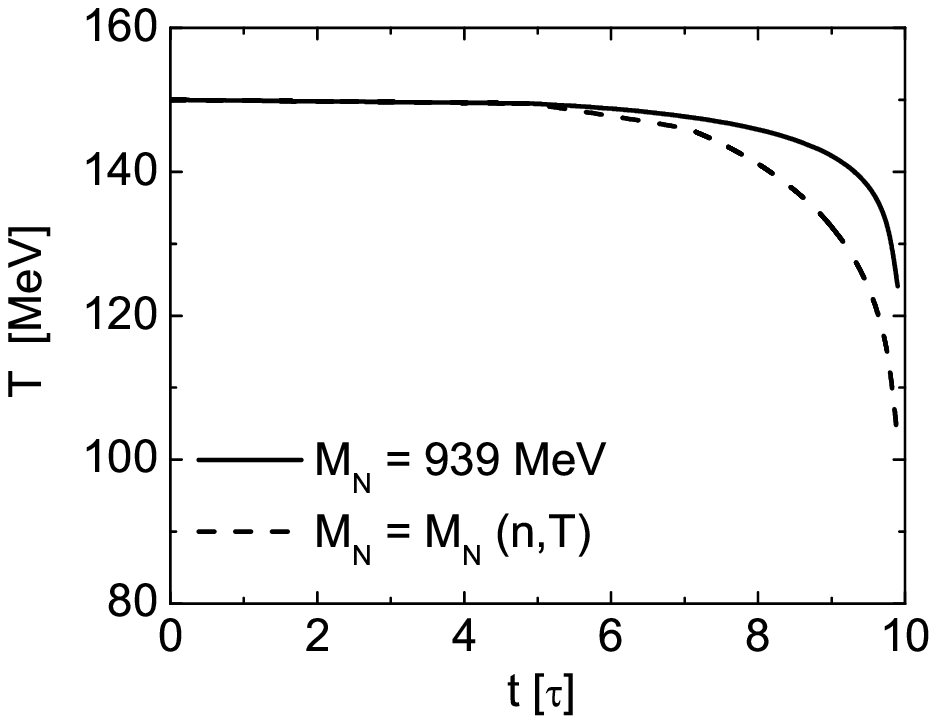}
\includegraphics[angle=0,scale=0.7]{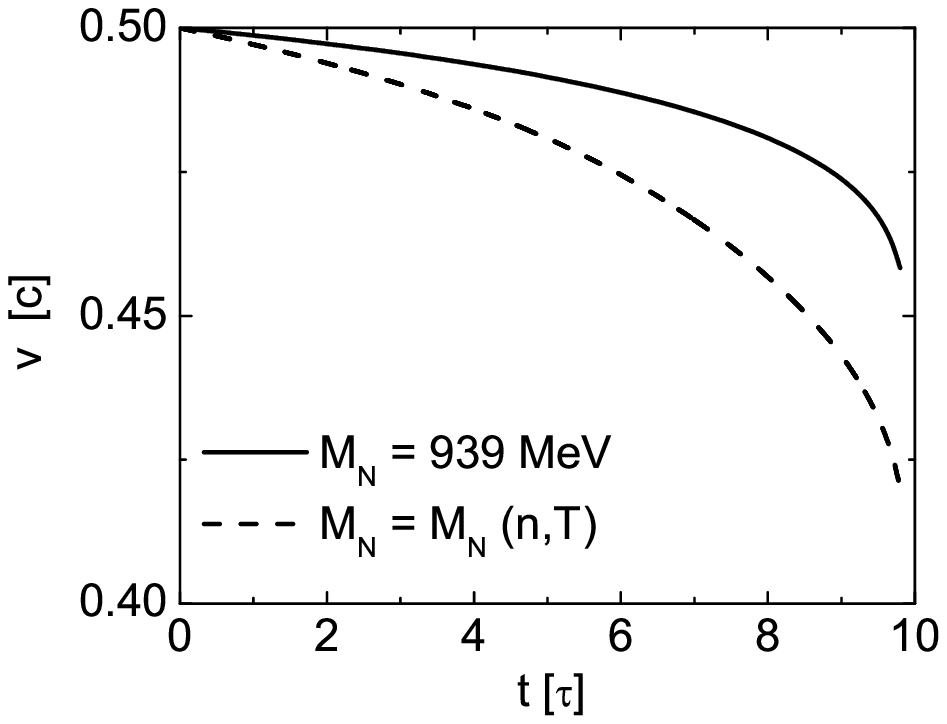}
\caption{
\bl: The temperature of the interacting component. 
\br: The flow velocity parameter $v$ of the interacting component.
The solid lines are with a constant nucleon mass $M_N (0)=939$ MeV, while 
the dashed curves are evaluated with a density and temperature dependent  
nucleon mass $M_N (n,T)$.\label{fig:tv}} 
\end{figure}
\end{center}

\begin{center}
\begin{figure}[!h]
\includegraphics[angle=0,scale=0.7]{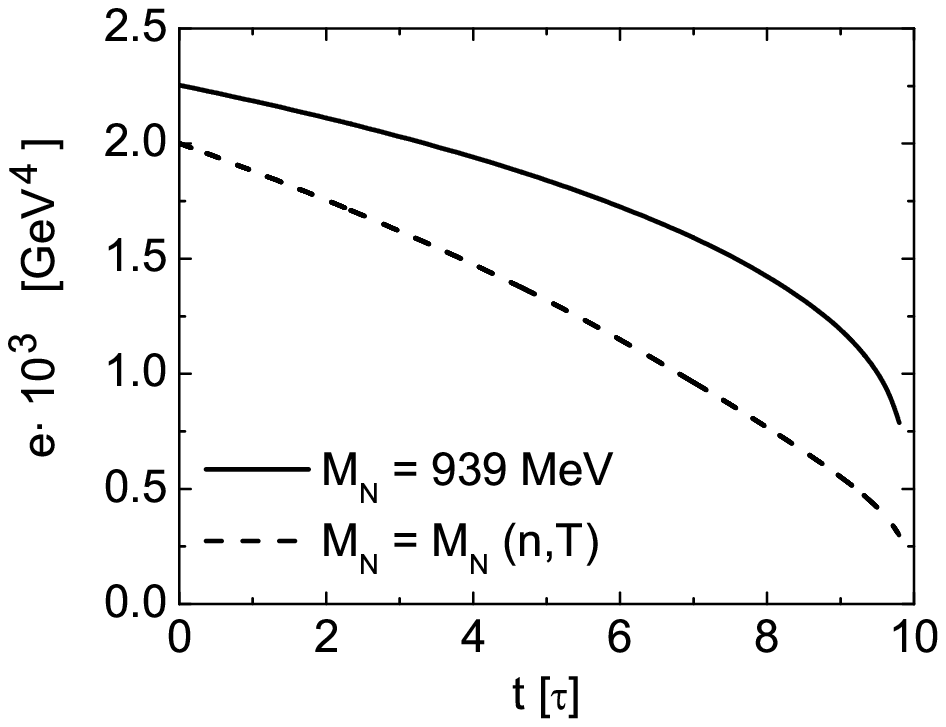}
\includegraphics[angle=0,scale=0.7]{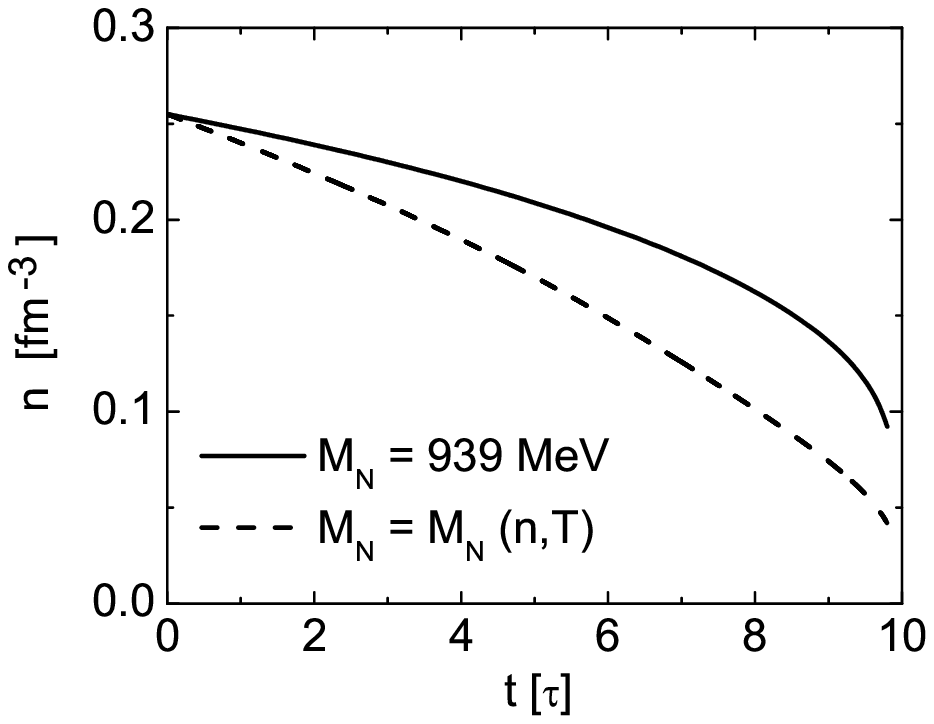}
\caption{
\bl: Nucleon energy density of the interacting component.
\br: Nucleon particle density of the interacting component. 
The solid lines are with a constant nucleon mass $M_N(0)=939$ MeV, while 
the dashed curves are evaluated with a density and temperature  
dependent nucleon mass $M_N (n,T)$. \label{fig:en} } 
\end{figure}
\end{center}

In Figs.~\ref{fig:tv} and \ref{fig:en}, the time evolution of the primary 
thermodynamical functions through the finite freeze out layer are shown,  
in terms of the proper time $\tau$. Note that the densities    
$n=n_i+n_f$ and $e=e_i+e_f$ are kept constant inside the layer. 

\begin{center}
\begin{figure}[!t]
\includegraphics[angle=0,scale=0.7]{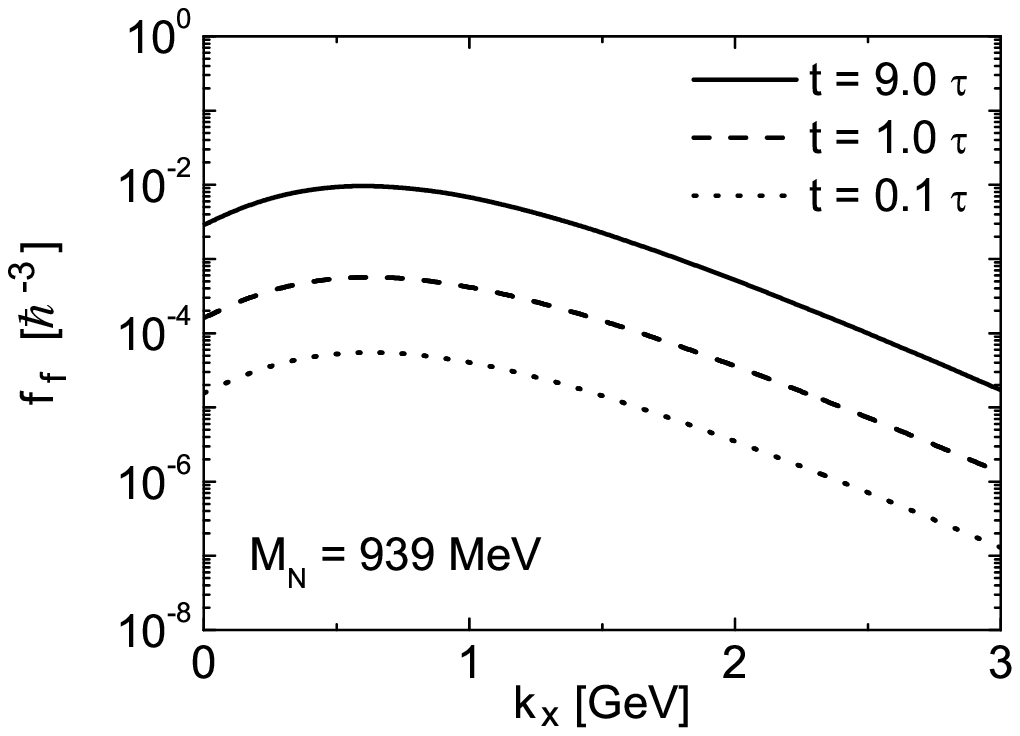}
\includegraphics[angle=0,scale=0.7]{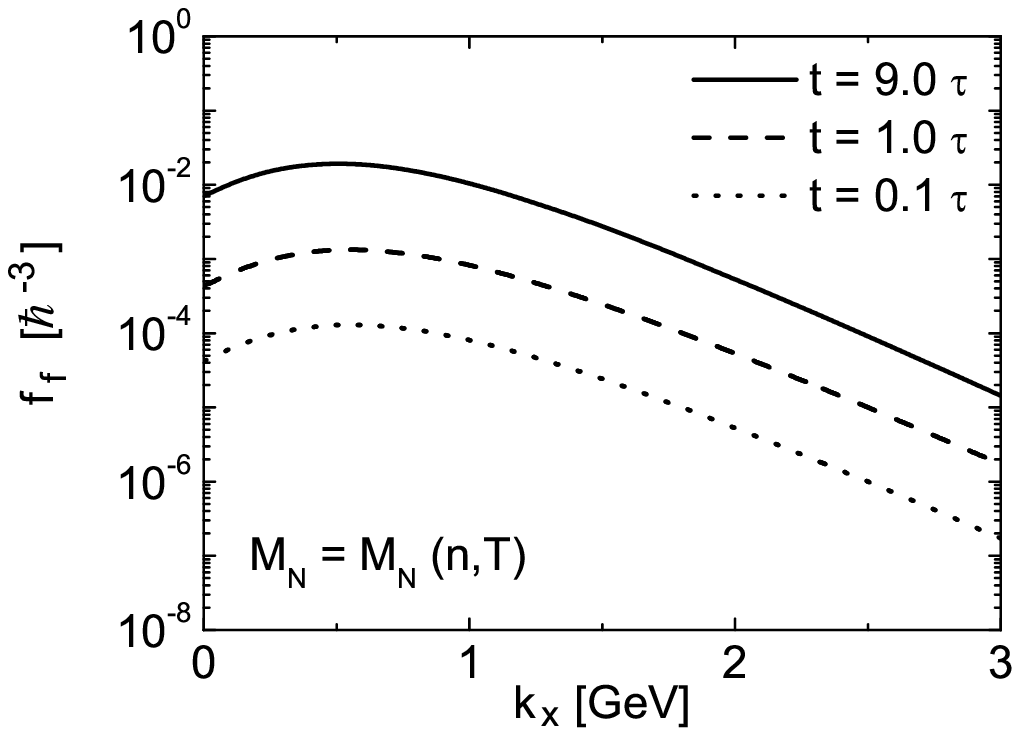}
\caption{\bl: Freeze out distribution function $f_f (k_x, k_y=0)$ at 
different instants $t$, evaluated
with a constant nucleon mass $M_N(0)=939$ MeV. 
\br: Freeze out distribution function $f_f (k_x, k_y=0)$ at different instants 
$t$, evaluated with the density and temperature 
dependent nucleon mass $M_N (n, T)$. The lines are as
in the left panel. The Freeze out distribution function 
has increased by a factor of $\simeq 2$, when density and temperatur 
dependent nucleon masses were taken into account.}
\label{fig:distr}
\end{figure}
\end{center}
We find a substantial impact of in-medium mass modification
on the freeze out process within the purely nucleon gas model.
Furthermore, the Figs.~\ref{fig:tv} and \ref{fig:en}
also elucidate, that the freeze out process
proceeds faster for all thermodynamical quantities $T, v, e, n$
when taking into account the mass dropping of nucleons.
The physical reason for a faster freeze out originates from a smaller
energy density of the nucleon system due to a smaller nucleon mass
$M_N(n,T)$ compared to the vacuum nucleon mass $M_N(0)$.

The given functions for $T, v, e, n$ are not directly accessible.
In experiments the way to study the hot and dense hadronic matter 
produced in heavy-ion collisions is to measure the distributions of final 
state particles, which reach the detectors long time after their 
last interaction. 
Accordingly, as next we consider the one-particle freeze out distribution 
function at $k_y=0$, i.e. $f_f (k_x) \equiv f_f (k_x, k_y=0)$ and consider 
the impact of the evaluated thermodynamical functions $T, v, e, n$ on it.
The results are shown in Fig.~\ref{fig:distr} 
for different instants during the freeze out process. 
The function $f_f (k_x)$ is determined at the point A \cite{footnote_A}
of the freeze out front; see also Ref. \cite{timelike_5} for more details. 
The function $f_f (k_x)$ is obtained 
by solving the differential equation  
(\ref{kinetic_15}), where for $f_i$ the J\"uttner distribution 
(\ref{Juttner_5}) is used, but with the parameters $T$ and $v$ as 
determined previously and given in Fig.~\ref{fig:tv}.

\begin{center}
\begin{figure}[!t]
\includegraphics[angle=0,scale=0.7]{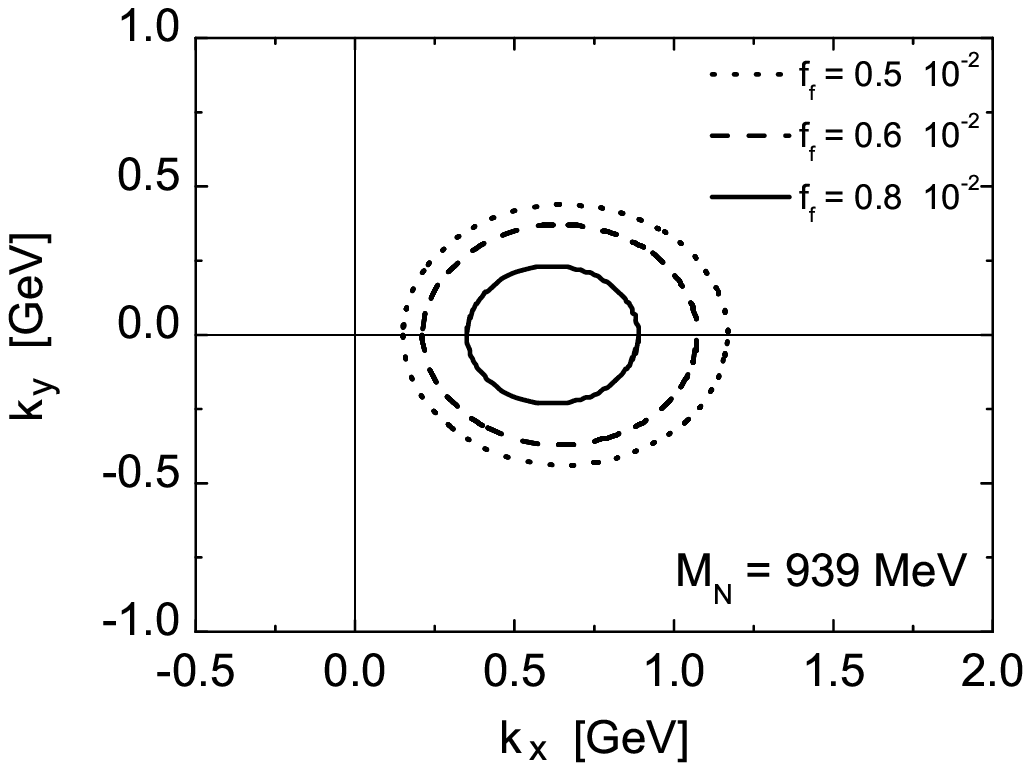}
\includegraphics[angle=0,scale=0.7]{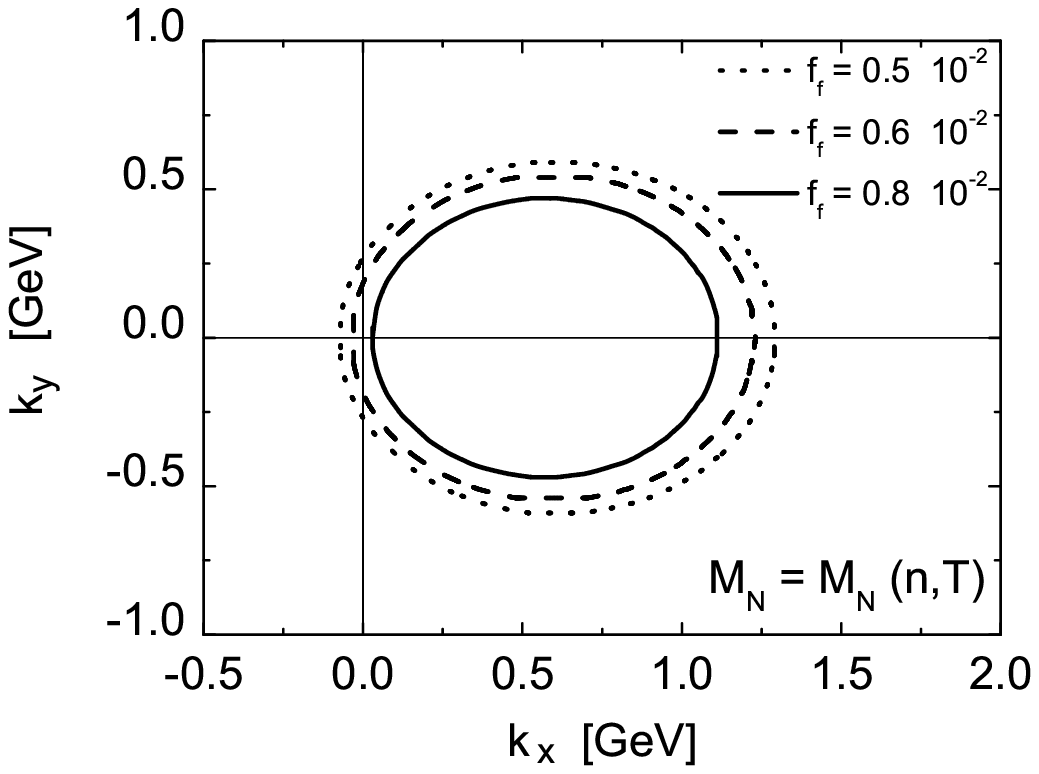}
\caption{Freeze out distribution functions $f_f (k_x, k_y)$ over 
their transversal and longitudinal momenta $k_x$ and $k_y$, respectively, 
at $t=9.0 \tau$ in RFF. 
\bl: Evaluation with a vacuum nucleon mass $M_N(0)=939$ MeV.
\br: Evaluation with the density and temperature dependent nucleon mass 
$M_N(n,T)$. 
The overall norm of the Freeze out distribution function 
has increased by a factor of $\simeq 4$ when density and temperature  
dependent nucleon masses were taken into account.} 
\label{fig:fig_4}
\end{figure}
\end{center}

The logarithmic scale in Fig.~\ref{fig:distr}
disguised the strong modification of these distribution functions. 
For small moments up to $k_x \le 1$ GeV, 
at the very beginning of the freeze out process at 
$t = 0.1 \, \tau$ there is a change of $f_f (k_x)$, 
which remains up to the end of the freeze out process 
at $t = 9.0\, \tau$.  
A contour plot of the freeze out particle distribution functions 
$f_f (k_x, k_y)$ over their transversal and longitudinal momenta 
$k_x$ and $k_y$, respectively,  
shown in Fig.~\ref{fig:fig_4}, illustrates this statement. 
We observe a remarkable change by a factor $\simeq 2$ 
for momenta $k_x, k_y \le $ 1 GeV.

A few remarks are in order about the used  
starting values for density and temperature.  
First, the formulas 
(\ref{massshift_4}) and (\ref{massshift_5}) have,  
like other theoretical approaches,
a limited range of validity in respect to the density; 
$n \le 1.5 \, n_0$.  
And second, according to Eq.~(\ref{kinetic_10}) the 
re-thermalization is assumed within a time step $d t$.   
Numerical accuracy for solving the set of differential equations  
(\ref{hydro_35}) - (\ref{hydro_45}) requires sufficiently  
small time intervalls $d t$. However, a smaller starting temperature 
$T_{{\rm pre}\;{\rm FO}}$ 
implies a longer re-thermalization time $\tau_0 < d t$, so that 
$T_{{\rm pre}\;{\rm FO}}$ cannot be taken arbitrary small.  
In addition, these two boundaries have to be adjusted to be  
in a region of the QCD phase diagram where we are inside the hadronization 
region and above the kinetic freeze out. 
The parameter choice of the starting values,  
$n_{{\rm pre}\;{\rm FO}} = 1.5 \, n_0$, $T_{{\rm pre}\;{\rm FO}} = 150$ MeV,  
are an optimal compromise for these borderlines.   
Within the approach presented we have a common way to model the kinetic 
freeze out process, and which is capable to implement the nucleon mass shift 
by means of a purely nucleon gas model.  
Nevertheless, one has also to be aware that the 
pion-nucleon ratio becomes small only for high enough nucleon densities
$n = (1.5 - 2) n_0$ and moderate temperatures $T \simeq 100$ MeV,  
e.g. \cite{becattini1}.
Our starting values for density and temperature on the pre freeze out 
hypersurface 
deviate from these values. Therefore, a more  
sophisticated model requires the implementation of pions and maybe even  
heavier mesons.  
However, due to different freeze out scenarios 
between nucleons and mesons, cf. \cite{Bratkovskaya}, 
such a procedure would require the use of a  
two-fluid or even three-fluid model, which is a highly involved tool, cf. 
\cite{two_fluid}. 
Therefore, for the
time being it is difficult to say how strong the impact of mesons is. 
Therefore, we were aiming at a description, which allows to account 
for the nucleon mass shift scenario during the freeze out process in 
a more common way. 

Finally, we remark that in-medium modifications have actually    
to be taken into account already before and during the hadronization process.
This points to an even 
stronger impact of in-medium modifications on the final particle
spectrum than presented. 

\section{Summary}\label{sum}

We have investigated a freeze out scenario within a finite layer 
for a massive nucleon gas. Special attention has been drawn about  
how strong the impact of the in-medium nucleon mass modification on the 
thermal freeze out process is. 
By focussing on a purely nucleon gas we have found a  
substantial effect 
on the thermodynamical quantities like 
temperature $T$, flow velocity $v$, 
particle density $n$ and 
energy density $e$  of the interacting component.  
All of these thermodynamical functions have revealed a faster freeze out 
compared to a scenario without in-medium nucleon mass shift.  
These modifications have a sizable implication on the 
freeze out particle distribution function, which is a basic 
observable in heavy-ion collision experiments.  
For small momenta around the nucleon mass a strong change of about 
a factor $\simeq 2$ has been found (see Fig.~\ref{fig:distr}). 
A contour plot of the particle distribution function in the 
transversal-longitudinal momentum plane $(k_x,k_y)$ illustrates this effect 
(see Fig.~\ref{fig:fig_4}). 
From these results we conclude  
that in-medium modifications of nucleons have a significant consequence on  
the freeze out process. This reasoning is certainly 
valid for heavy-ion collisions, which produce sufficiently high 
nucleon particle densities; in particular for experiments like 
Compressed Baryonic Matter (CBM collaboration)  
planned at the GSI facility in Darmstadt/Germany.

For a more realistic description of heavy ion collisions 
one should include in the analysis at least the low laying mesons 
and baryons as well.
All hadrons suffer in-medium modifications of their masses and widths,  
but there are strong differences among them. For instance, while  
the pion mass remains almost unaffected by the hadronic medium 
even at very high temperatures 
and densities, this is not the case for nucleons, kaons and Delta resonances.
Taking into account the pions and the in-medium modifications of other 
hadrons in the fireball produced in nucleus-nucleus reactions could modify 
our results in the details, but not the general statement that 
in-medium modifications have some relevance for the freeze out process. 
For example, the implementation of the pions leads to a 
stronger temperature dependence of the chiral condensate 
\cite{change_condensate2}, which causes a stronger down shift of the 
nucleon mass with increasing temperature.  
Then, our results might even be more pronounced.  
In addition, the implementation of in-medium modifications has to be 
taken into account before and during the hadronization, which leads 
also to an amplification of their impact on the whole freeze out process.  

In summary, our findings for a purely nucleon gas suggest 
that taking into account in-medium modifications of nucleons 
seems to be a necessary and interesting phenomenon,  
in particular for collision scenarios with high baryonic densities. 

% ========================================================================
% Acknowledgements
% ========================================================================
\section*{Acknowledgements}  
The authors would like to express their gratitude to 
Prof. Jean Cleymans for enlighting discussions. 
S.Z. thanks for the warm hospitality at the Bergen Center for Computational 
Science (BCCS) and Bergen Physics Laboratory (BCPL) at the University of 
Bergen/Norway. 
S.Z. and J.M. wishes to acknowledge the NordForsk for the partial financial
support of this work.
% ========================================================================
% APPENDIX
% ========================================================================

\section*{Appendix}

The function $G_n^{\pm}\; (n=1,2)$ are defined as 
\bea
G_n^{\pm} (M_N, v, T) &=& \frac{1}{T^{n+2}} \, \int\limits_0^{\infty} 
d k \; k \left( \sqrt{k^2 + M_N^2} \right)^n 
\nonumber\\
&&\times {\rm E}_1  
\left(\frac{\gamma}{T} \, \sqrt{k^2 + M_N^2} \pm \frac{\gamma\, v\, k} 
{T}\right)\;, 
\nonumber\\
\label{appendix_5}
\eea 
where ${\rm E}_1$ is a special case of incomplete Gamma-function 
\cite{Abramowitz_Stegun} and defined as  
\bea
{\rm E}_1 (x) &=& \int \limits_{x}^{\infty} dt \; t^{-1} \; 
{\rm e}^{- t}\;. 
\label{appendix_10}
\eea 
The function $K_n$ is the Bessel function of second kind 
\cite{Abramowitz_Stegun}, defined as  
\bea
K_n (z) = \frac{2^n\, n!}{(2 n)!} z^{-n} \int_x^{\infty} dx 
{\rm e}^{-x} \, (x^2 - z^2)^{n - 1/2}\,.
\label{appendix_15}
\eea
Finally, we prove the vanishing of the second term in Eq.~(\ref{hydro_35}). 
First, we note explicitly the relevant four-current and 
energy-momentum tensor components as deduced directly from the 
microscopic kinetic definitions (\ref{hydro_15}) and (\ref{hydro_10}), 
respectively. 
We recall that due to the immediate re-thermalization limit 
during the freeze out there is actually a J\"uttner type distribution 
for $f_i$, but with the thermodynamical functions $T$ and $v$ as evaluated 
with the approach presented and given in Figs.~\ref{fig:tv}. 
Therefore, at the beginning of the time-step 
for $f_i$ one has to insert the J\"uttner distribution 
(\ref{Juttner_5}), but with the evaluated functions $T$, and $v$,  
into the microscopic definitions (\ref{hydro_15}) 
and (\ref{hydro_10}), getting the following components in RFF:  
\bea
N^0 &=& \frac{n}{4} \left[2 \, a \, b\, K_0 (a) + 4 \, b\, K_1 (a)\right]\;,
\label{prove_5} 
\eea
\bea
N^x &=& \frac{n}{4} \left[2 \, v \, a\, b\, K_0 (a) + 
4 \, v \, b \, K_1(a) \right] \;,
\label{prove_10} 
\eea
\bea
T^{0 0} &=& \frac{n T}{4} \left[2 a \, b^2 \, K_1 (a) + 
2 \, b^2 \, (3 + v^2) K_2 (a) \right]\;,
\nonumber\\
\label{prove_15} 
\eea
\bea
T^{0 x} &=& - \frac{T}{\gamma\, v} N^0  
\nonumber\\
&& + \frac{n T}{4} \left[2 a \, b^2 \, v \, K_1 (a) + 
2 \frac{b^2}{v} (1 + 3 v^2) K_2 (a) \right]\;,
\nonumber\\
\label{prove_20} 
\eea
\bea
T^{x x} &=& - 2 \frac{T}{\gamma \, v} N^x 
\nonumber\\
&& + \frac{n T}{4} 
\bigg[2 \, a\,b^2 \, v^2 \, K_1 (a) + 2 \, b^2 (3 + v^2) K_2(a) \bigg]\;.
\nonumber\\
\label{prove_25}
\eea
We recall that $a = M/T$, and $b= \gamma a$ with $\gamma = (1 - v^2)^{-1/2}$. 
The second term in Eq. (\ref{hydro_35}) is given as 
\bea
d u_{\mu} \, T^{\mu \nu}\, u_{\nu} &=& d u_0 \, T^{0 0} u_o 
+ d u_0 \, T^{0 x} u_x 
\nonumber\\
&& + \, d u_x  T^{x 0} u_0 
+ d u_x \, T^{x x} u_x\;. 
\label{prove_30}
\eea 
With $u_{\mu} = \gamma (1, -v, 0, 0)$ we get 
$d u_0 = \gamma^3 \, v \, d v$ and 
$d u_x = - \gamma^3 \, d v$.
By using these relations and inserting the 
components (\ref{prove_5}) - (\ref{prove_25}) into (\ref{prove_30}) 
we immediately find $d u_{\mu} T^{\mu \nu} u_{\nu} = 0$.  
We recall that $a K_2 (a) = a K_0 (a) + 2 K_1 (a)$.


\begin{thebibliography}{99}
\bibitem{GellMann_Oakes_Renner} When a numerical value for the
chiral condensate is fixed, then the current quark mass
$m_q = (m_u + m_d)/2$ is determined by the
Gell-Mann--Oakes--Renner relation: $m_{\pi}^2 f_{\pi}^2 = - 2 m_q
\langle 0| {\overline q} q | 0\rangle$; $f_{\pi} = 92$ MeV is the pion decay
constant and $m_{\pi}= 138$ MeV is the pion mass.
\end{thebibliography}
\end{document}